\documentclass[format=acmsmall, review=false]{acmart}
\makeatletter
\def\@ACM@checkaffil{
    \if@ACM@instpresent\else
    \ClassWarningNoLine{\@classname}{No institution present for an affiliation}%
    \fi
    \if@ACM@citypresent\else
    \ClassWarningNoLine{\@classname}{No city present for an affiliation}%
    \fi
    \if@ACM@countrypresent\else
        \ClassWarningNoLine{\@classname}{No country present for an affiliation}%
    \fi
}
\makeatother
\usepackage{acm-ec-23}
\usepackage{booktabs} 
\usepackage[ruled]{algorithm2e} 

\SetAlFnt{\small}
\SetAlCapFnt{\small}
\SetAlCapNameFnt{\small}
\SetAlCapHSkip{0pt}
\IncMargin{-\parindent}

\setcitestyle{authoryear}

\usepackage{amsmath,amsfonts}
\usepackage{mathtools}
\usepackage{cleveref}
\usepackage[export]{adjustbox}

\newcommand{\ap}[1]{{\color{red}[Ariel: #1]}}
\renewcommand{\dh}[1]{\textcolor{blue}{[Daniel: #1]}}
\newcommand{\is}[1]{\textcolor{teal}{[Itai: #1]}}

\newcommand{\emdash}{\,---\,}

\DeclarePairedDelimiter{\set}{\{}{\}}
\DeclarePairedDelimiter{\inner}{\langle}{\rangle}
\DeclarePairedDelimiter{\norm}{\|}{\|}
\newcommand{\eng}{\textsc{eng}}

\newcommand{\divers}{\textsc{div}}
\renewcommand{\b}{\mathbf{b}}
\newcommand{\A}{\mathbf{A}}
\newcommand{\p}{\mathbf{p}}
\newcommand{\R}{\mathbb{R}}
\newcommand{\I}{\mathbf{I}}
\newcommand{\x}{\mathbf{x}}
\newcommand{\y}{\mathbf{y}}
\newcommand{\e}{\mathbf{e}}
\renewcommand{\c}{\mathbf{c}}

\newcommand{\fol}{\textsc{following}}
\renewcommand{\d}{\delta}
\newcommand{\engopt}{OPT^{eng}}
\newcommand{\divopt}{OPT^\delta}
\newcommand{\dcost}{\textsc{cost}^\d}
\DeclareMathOperator*{\argmax}{arg\,max}
\newcommand{\inc}{\textsc{inc}}
\allowdisplaybreaks

\title{Optimal Engagement-Diversity Tradeoffs in Social Media}

\newif\ifanonymized
\anonymizedfalse
\ifanonymized
\author{Submission \#337}
\else
\author{Fabian Baumann}
\affiliation{\institution{Max Planck Institute for Human Development}}

\author{Daniel Halpern}
\affiliation{\institution{Harvard University}}

\author{Ariel D. Procaccia}
\affiliation{\institution{Harvard University}}

\author{Iyad Rahwan}
\affiliation{\institution{Max Planck Institute for Human Development}}

\author{Itai Shapira}
\affiliation{\institution{Harvard University}}

\author{Manuel W\"uthrich}
\affiliation{\institution{Harvard University}}

\fi
\begin{abstract}
Social media platforms are known to optimize user engagement with the help of algorithms. It is widely understood that this practice gives rise to echo chambers\emdash users are mainly exposed to opinions that are similar to their own. In this paper, we ask whether echo chambers are an inevitable result of high engagement; we address this question in a novel model. Our main theoretical results establish bounds on the maximum engagement achievable under a diversity constraint, for suitable measures of engagement and diversity; we can therefore quantify the worst-case tradeoff between these two objectives. Our empirical results, based on real data from Twitter, chart the Pareto frontier of the engagement-diversity tradeoff. 
\end{abstract}

\begin{document}

\begin{titlepage}

\maketitle

\end{titlepage}

\section{Introduction}\label{sec:intro}

It is no secret that social media companies heavily rely on algorithms to optimize user engagement. This practice has a well-documented dark side that is widely scrutinized and debated. For example, writing recently in the New York Times, the technology pioneer Jarron Lanier coins the term ``Twitter poisoning'' to describe ``a side effect that appears when people are acting under an algorithmic system that is designed to engage them to the max'' \citep{Lan22}.

Perhaps the main reason that optimized engagement is so broadly decried is that it may lead to increased political polarization through the formation of echo chambers, where users are only exposed to viewpoints and opinions that closely align with their own. In the pages of the Washington Post, \cite{JKG20} lament that ``the features that facilitate a right-wing echo chamber on Facebook\emdash such as [\ldots] how the algorithms work to maximize engagement\emdash are intentional choices.''\footnote{Ironically, they give Twitter as a positive example, at least in terms of the stated intentions of its founder and former CEO, Jack Dorsey.} This statement is supported by a paper by the same authors~\citep{KJG20} and grounded in a large, important body of prior work~\citep{Stroud10,Par11,CRFG+11,Suns18}.

In several studies, however, there is somewhat mixed evidence for the relation between algorithms, the diversity of content users consume, and political polarization~\citep{BMA15,FGR16}. In fact, an influential paper casts doubt on the very idea that eliminating echo chambers and showing more diverse information sources leads to reduced polarization~\citep{BABB+18}. But even the lead author of that paper concedes that, on Twitter, ``breaking up the echo chambers that prevent cross-party discussion about market-based solutions to climate change, for example, might be more successful'' than having broad conversations about politics \citep{Bail18}.

This debate notwithstanding, it seems that academics and pundits largely agree on one underlying assumption: There is a tradeoff between user engagement and the diversity of information they are exposed to. In other words, if a social media platform wishes to maximize engagement and optimize its revenue, it would necessarily have to expose users to the posts or tweets they are most likely to engage with, thereby limiting diversity of information and creating echo chambers. 

In this paper, we aim to \emph{quantify} the engagement-diversity tradeoff. Our high-level research question is this: 
\begin{quote}
	\emph{How much engagement must be sacrificed in order to guarantee a given level of diversity of information?}
\end{quote}
We are particularly interested in identifying scenarios where diversity of information comes at little cost to engagement, as in such scenarios, it is more likely that social media platforms would be willing to break up echo chambers. 

\subsection{Our Approach and Results}

We base our terminology on Twitter, but our model and analytical results are relevant to most social media platforms, including Facebook. As usual, we represent the social network as a directed graph,\footnote{For Facebook we would simply have bidirectional edges between friends.} where the nodes are users and an edge from $i$ to $j$ means that $i$ follows $j$. We also assume that tweets are partitioned into $T$ types, and each user $i$ has probability $p_{ti}$ of retweeting a tweet of type $t$.

We model the propagation of tweets in the network as a discrete-time (Markov decision) process where, in each step, users are exposed to tweets from their followees, as well as tweets directly shown to them by the platform. This latter component is the algorithmic \emph{injection policy}, which presents to each user tweets of different types, subject to a budget constraint. Note that this modeling choice matches Twitter's default view, where users are shown a combination of tweets shared by their followees and some algorithmically selected tweets that originate in non-followees. The injection policy itself is time-independent, that is, users are exposed to the same mixture of types in each round; as we prove, this is without loss of generality (assuming that the network and retweet probabilities are fixed, of course). 

We can now quantify \emph{engagement} by measuring the number of retweets in the system (in the limit as the number of rounds grows). This is a natural measure in our simple model; a more elaborate model may take into account the affinity of users to different tweet types; as we discuss in Section~\ref{sec:disc}, our results extend to this setting. There are many ways to quantify \emph{diversity} in our model; our measure is the minimum, across users $i$ and tweet types $t$, of the number of tweets of type $t$ seen by user $i$ (in the limit as the number of rounds grows). This is a rather onerous choice, as it requires that every user be exposed to every tweet type; such a pessimistic view means that any positive results are especially robust. 

To analyze the tradeoff between engagement and diversity, we are interested in two injection policies: the one that maximizes engagement and the one that maximizes engagement subject to achieving at least $\delta$-diversity for a given $\delta\geq 0$. The \emph{cost of $\delta$-diversity}, then, is the fraction of the engagement of the former policy that is sacrificed by employing the latter policy. 

Our main theoretical result is an upper bound on the cost of $\delta$-diversity. Assuming the average retweet probability of each user is at least $\alpha$ and each retweet probability is at most $\beta>0$, and that $\delta\leq 1/T$, the cost of $\delta$-diversity is at most $T\delta(1-\alpha/\beta)$, and this bound is tight. Qualitatively, the implication is that with a user base that is generally engaged (high average retweet probability $\alpha$ compared to the maximum $\beta$), the cost of $\delta$-diversity is small. This bound is both trivial and tight in the special case where the graph is empty; our result is encouraging in that it demonstrates that the very same bound still holds despite the nontrivial complications arising from the dynamics of retweets in a general social network. 

To obtain a more nuanced understanding of the engagement-diversity tradeoff in practice, we also conduct experiments on a large dataset from Twitter. We process the data to extract the social network graph and, based on hashtags, infer four types of tweets and their associated retweet probabilities. We then measure the cost of $\delta$-diversity as the retweet probabilities are scaled up. The results show that the practical tradeoff is far better than the worst-case bound, and that the cost of diversity is typically (though, surprisingly, not always) monotonically decreasing in the magnitude of retweet probabilities. Finally, we discuss how a policy-maker can operationalize these results.

\subsection{Related Work}

Needless to say, the literature on social networks and recommender systems is vast. Here we elaborate on a few recent papers in this area. 

Several papers~\cite{LH19,AMMA+20,HCCN+20} consider the impact of algorithms on diversity through field experiments. In particular, in work presented at EC a few years ago, \citet{HCCN+20} study the engagement-diversity tradeoff via a field experiment on Spotify. Their control and treatment groups were given podcast recommendations to maximize engagement; in the case of the treatment group, the recommendation algorithm was personalized, whereas, in the case of the control group, recommendations were based on demographics. Treatment significantly increased engagement and significantly decreased diversity (measured through the category tags of podcasts). The authors conclude that ``these findings highlight the need for academics and practitioners to continue investing in personalization methods that explicitly take into account the diversity of content recommended.''

Similarly, \citet{HKOB+21} report results from a field experiment on Twitter, where the control group was shown tweets in reverse chronological order, without algorithmic personalization. They find evidence for algorithmic amplification of certain political groups; specifically, they conclude that the mainstream political right enjoys higher algorithmic amplification than the mainstream political left. This paper reinforces the connection between algorithms and political polarization on Twitter, but it does not examine the engagement-diversity tradeoff. 

\citet{SBMR22} apply machine learning to a large set of Twitter data in order to predict the relation between the text of a tweet and its ability to engage a politically diverse audience. They further incorporate this predictor into a tool that crafts tweets that are appealing to users across the political spectrum. This intriguing approach is beyond the scope of our model, as we treat categories and retweet probabilities as fixed.

\section{Model and Machinery}\label{sec:model}
In this section, we introduce our stylized model of the Twitter social network, its dynamics, and key definitions. Additionally, we describe a computational framework for analyzing the engagement-diversity tradeoff and provide theoretical results in support of the robustness of our modeling choices. 

\subsection{Definitions}\label{subsec:definitions}

\paragraph{Social network instance} There is a set of $n$ \emph{users} denoted $[n] = \set{1, \ldots, n}$. As is standard in social networks, users may follow each other. We represent this in the canonical way using a \emph{follower graph} $G = ([n], E)$ with users as vertices and a (directed) edge $(i, j) \in E$ present when user $i$ follows user $j$.  We use $\fol(i)$ to denote the number of users $i$ follows (i.e., $i$'s outdegree in $G$).  

There are $T$ \emph{types} of tweets $[T] = \set{1, \ldots, T}$ (indexed by $t$) users can view in their feeds of content. Users' feeds are determined by what type of tweets the users they follow engage with. By retweeting, the user propagates this type through the follower graph. In the next timestep, followers of the user will be able to view and further distribute the tweets. Thus, specific types of content can spread through the network from a small set of initial adopters to a potentially much larger group.

In more detail, it is assumed that upon seeing a tweet, each user has a probability of retweeting this tweet determined by its type. This is denoted by a type's \emph{retweet probability vector} $\p_t \in [0, 1)^{n}$ where $p_{ti}$ is the probability user $i$ retweets a tweet of type $t$. We use $\p = (\p_t)_{t \in [T]}$ to denote the collection of retweet probability vectors for all types. We assume that followers do not get the exact same tweet
as the original tweet, in order to avoid addressing the case of a retweet loop in this theoretical framework. Instead,
retweeting propagates the type and not its specific realization.

\paragraph{States} A \emph{type state} $\x^{(k)}_t \in \R^{n}$ represents the expected number of tweets of type $t$ being seen by the users. The component $x^{(k)}_{ti}$ is the (expected) number of tweets of type $t$ seen by a user $i$ at time $k$. We use $\x^{(k)} = (\x^{(k)}_t)_{t \in [T]}$ to denote the collection of all type states and simply call it a \emph{state}. We will use $\x$ to refer to an unparameterized state (without a timestep). 

\paragraph{User feed} In our model, there are two ways a tweet may end up in a user's state (i.e., be seen on their feed), either directly from another user they follow or \emph{injected} by the social network. In general, we will assume that the former is exogenous and given as part of the system while the latter is a policy we (as the social network) have control over. 
To understand the former, suppose a user $i$ follows a user $j$ and $j$ sees a tweet of type $t$ at time $k$.
The probability that $i$ sees this tweet at time $k + 1$ we assume to be $\frac{p_{tj}}{\fol(i)}$, that is, it is the probability that $j$ retweets this tweet scaled down by the number of users $i$ follows. Observe that this is well-defined as $\fol(i) \ge 1$ by the assumption that $i$ follows $j$. The scaling down is to account for the fact that if $i$ follows many users, they will not necessarily see all the retweets in their feed.\footnote{This scaling down is admittedly a controversial modeling choice. One could alternatively assume that each user has a fixed ``attention budget,'' but that would lead to nonlinear dynamics. Scaling down by $\fol(i)$ is a way of realistically bounding the number of tweets users see\emdash after all, if a user follows thousands of people, they will not have time to peruse all their tweets\emdash while preserving the linearity of the model. It thus strikes a good balance between realism and technical tractability.}  
 We use a \emph{type matrix} $\A_t \in \mathbb{R}^{n \times n}$ to store these seen probabilities, where
$$
A_{tij} = \begin{cases} \frac{p_{tj}}{\fol(i)} & \text{if } i \text{ follows } j\\
0 & \text{otherwise}
\end{cases}
$$
We again use $\A = (\A_t)_{t \in [T]}$ to denote the collection of all type matrices. Note that if the state for type $t$ at time $k$ is $\x^{(k)}_t$, then at time $k + 1$, each user will see $\A_t \x^{(k)}_t$ based on retweets only. 

To represent a social network's injection, we define an injection policy $\b = (\b_t)_{t \in [T]}$ where each $\b_t \in [0, 1]^n$. The component $b_{ti}$ represents the number of expected number tweets shown to user $i$ of type $t$.  In addition, we require that for each user $i$, $\sum_t b_{ti} \le 1$, i.e., only at most one tweet ``unit''  can be injected at each time step. 

\paragraph{Dynamics and limiting behavior} For an injection policy $\b$, we obtain the following dynamics on tweets seen in the system. At time $0$ for each type $t$, we simply have $\x^{(0)}_t = \b_t$. For all times $k \ge 0$, we have $\x^{(k + 1)}_t = \A_t\x_t^{(k)} + \b_t$; in words, the tweets seen by users in time $k + 1$ are retweets by others they follow along with direct injections to them. Abusing notation slightly, we allow matrix and vector operations to work over collections, i.e., writing $\x^{(k + 1)} = \A\x^{(k)} + \b$ to refer to all types. Notice that for each type individually, this is a standard linear dynamical system. However, this formulation is unusual because the constraint on policies $\b$ (one unit per user) is across types, creating interdependences. 

Unraveling the recursion, we see that the $k$'th timestep can be written as $\x^{(k)} = \sum_{\ell = 0}^{k - 1} \A^\ell \b$ and, by linearity, it can be written as $\x^{(k)} = (\sum_{\ell = 0}^{k - 1} \A^\ell)\b$. Since the sum of each row of $\A_t$ is strictly less than one,
the spectral radius of each type, $\rho(\A_t)$, is less than one (this follows from, e.g., the Gershgorin circle theorem). This implies that
the limit $\lim_{k \to \infty} \sum_{\ell = 0}^{k - 1} \A^\ell$ exists and approaches $(\I - \A)^{-1}$, where $\I$ is the identity matrix~\citep{hubbard2015vector}. Since this limit matrix will come up so often, we will use the notation $\A^*_t = (\I - \A_t)^{-1}$ and similarly $\A^* = (\I - \A)^{-1}$. However, using this detail, we see that the state also converges to a ``limiting state'' $\A^*\b$. We will use $\x(\b) = \A^*\b$ to denote the limiting state of policy $\b$ (recall this really means the collection of $\x(\b)_t = (\A^*_t)^{-1}\b_t$). 

\paragraph{Engagement and diversity}
Two desirable properties guide our analysis. The \emph{engagement} of a state $\x$ is denoted $\eng(\x) = \sum_t \langle \p_t, \x_t \rangle$. It captures the expected number of retweets generated by the state $\x$. In some sense, this assumes that a user's ``engagement'' with a tweet is simply the likelihood they are to retweet it. However, as we discuss in \Cref{sec:disc}, the coupling of engagement and retweet probability is unnecessary; we primarily do so for ease of presentation as it seems like a reasonable choice for such an engagement parameter. The \emph{diversity} of a state $\x$ is denoted $\divers(\x) = \min_{t \in [T], i \in [n]} x_{ti}$, i.e., the fewest tweets of any type seen by any user. We say that a state \emph{satisfies $\d$-diversity} if $\divers(\x) \ge \d$. Additionally, we extend the notions of engagement and diversity to injection polices by simply having them operate on their limiting state. Formally, we have $\eng(\b) = \eng(\x(\b))$, $\divers(\b) = \divers(\x(\b))$, and $\b$ satisfies $\d$-diversity exactly when $\x(\b)$ does.

We let $\engopt(G, \p)$ be the optimal engagement for graph $G$ and retweet probabilities $\p$, that is, the maximum over injection policies $\b$ of $\eng(\b)$. We sometimes will write $\engopt$ if $G$ and $\p$ are clear from the context. 

To understand the engagement-diversity tradeoff, we are especially interested in the optimal engagement achievable under a diversity constraint. We denote this by $\divopt(G, \p)$, parameterized by $\d$, that is, the maximum over injection polices $\b$ with $\divers(\b)\geq \d$ of $\eng(\b)$. We will again sometimes write $\divopt$ if $G$ and $\p$ are clear from context. Notice that it is always feasible to guarantee $\d$-diversity for $\d \le 1/T$ since the policy with $b_{ti} = 1/T$ for all $i$ and $t$ achieves this. However, for $\d > 1/T$, there are instances where no policy achieving $\d$-diversity exists. Hence, from now on, we will only focus on $\d \le 1/T$.

\paragraph{Cost of $\d$-diversity} Finally, we define $\dcost(G, \p) = 1 - \frac{\divopt}{\engopt}$. This captures the multiplicative loss on optimal engagement by imposing $\d$-diversity, i.e., a cost of $.2$ for $\d = .1$ means that 20\% of engagement is lost by enforcing $.1$-diversity. From another perspective, $1 - \dcost(G, \p)$ as a function of $\d$ plots the Pareto frontier of the trade-off between engagement and a given diversity level. 

\subsection{Optimizing Engagement and Diversity}\label{subsec:optimizing}

As it turns out, computing $\engopt$ amounts to solving a linear program. Namely, we have that
\[
\eng(\b) = \sum_t \inner{\p_t, \x(\b)_t} = \sum_t \inner*{\p_t, \A^*_t \b_t} = \sum_t p_t^{\top}\A^*_t \b_t.
\]
This is a linear objective in variables $b_{ti}$ for $t \in [T]$ and $i \in [n]$. This objective will come up quite often throughout our analysis, so we introduce the notation $\c_t = (p_t^{\top}\A^*_t)^{\top}$ to be the vector of coefficients on the $\b_t = (b_{t1}, \ldots, b_{tn})$ variables. As before, we use the notation $\c = (\c_t)_{t \in [T]}$. We can interpret a value $c_{ti}$ as the total engagement generated in the system by injecting a unit of type $t$ to user $i$. This allows us to write the engagement as
\[
    \eng(\b) = \sum_t \c_t^{\top} \b_t.
\]

Since the constraints of being a valid injection policy are also linear, we can write the whole program as
\begin{alignat*}{3}
	 &\text{maximize:}  & &\sum_t \c_t^{\top} \b_t  && \\
	 &\text{subject to:}  \quad && \sum_t b_{ti} \le 1, \quad&&  i \in [n]\\
	 &&& b_{ti} \ge 0 \quad&& t \in [T], i \in [n].
\end{alignat*}

We will refer to this linear program as the \emph{engagement-optimal program}.
An interesting observation is that the optimal value and solutions of the program have a simple closed form. Notice that there are no constraints involving distinct users; the only constraint is that for each user, the total injection is at most one. Hence, the optimal policy is to spend this budget of one only on the tweet type $t$ with the largest objective coefficient $c_{ti}$. In other words, an optimal policy is to:
\begin{enumerate}
    \item for each user $i$, set a single $b_{ti}$ for a type $t$ maximizing $c_{ti}$ to $1$ (or any linear combination of maximizing types), and
    \item set all other $b_{t'i}$ to $0$.
\end{enumerate}
This achieves engagement $\sum_i \max_t c_{ti}$. 

Things become less straightforward if we wish to optimize engagement subject to $\d$-diversity, that is, if we wish to compute $\divopt$. The $\d$-diversity constraint is also linear, so this remains a linear program, as follows:
\begin{alignat*}{3}
	 &\text{maximize:}  & &\sum_t \c_t \b_t && \\
	 &\text{subject to:}  \quad && \sum_t b_{ti} \le 1, \quad &&  i \in [n]\\
	 &&& (\A^*_t\b_t)_i \ge \d, \quad && t \in [T], i \in [n]\\
	 &&& b_{ti} \ge 0 \quad && t \in [T], i \in [n].
\end{alignat*}
We refer to this program as the \emph{$\d$-diversity program}.
Unlike before, however, it does not seem to have a concise closed form. 

\subsection{Robustness of the Modeling Choices}\label{sec:limit} 
One modeling choice that may initially seem unnatural is to define engagement and diversity in the limiting state. We did this as it led to cleaner statements of our results and more straightforward experiments. However, this section aims to show that other reasonable choices lead to essentially the same model in terms of engagement-diversity trade-offs, justifying our choices.

An alternate formulation is to have a time horizon $K$ and consider what occurs at each timestep. For example, one could define engagement to be the average (or equivalently sum) engagement over all timesteps, i.e., $\frac{1}{K + 1}\sum_{k = 0}^K\eng(\x^{(k)})$. Similarly, one could instead require that $\d$-diversity be satisfied at \emph{every} timestep rather than just in the limit. In such a model, requiring injection policies to be identical in every timestep may be overly restrictive. A priori, it seems plausible that substantially better policies exist that change over time; for example, they can oscillate between different injections or modify what they inject once certain levels of diversity have spread through the network. Hence, we could even allow injection policies to be \emph{time-dependent}, changing what they inject depending on the timestep (in contrast to our \emph{time-independent} program solutions). \Cref{thm:app-optimal} suggests that these decisions do not fundamentally impact the model. Perhaps surprisingly, the time-independent engagement policies computed by our programs remain approximately optimal with these alternative definitions, which holds \emph{even when compared to the more powerful time-dependent strategies}. We therefore expect such a model to lead to qualitatively similar results.

\begin{theorem}\label{thm:app-optimal}
    Fix $\d$ and let $\b^*$ be a solution to the $\d$-diversity program. Fix a time horizon $K$ and let $\x^{(0)}, \ldots, \x^{(K)}$ be the states induced by injecting $\b^*$ at every timestep. Using the notation $\eng^{av}\left(\x^{(0)}, \ldots, \x^{(K)}\right) = \frac{1}{K + 1}\sum_{k = 0}^K \eng(\x^{(k)})$, we then have:
    \begin{enumerate}
        \item The notions of engagement on $\b^*$ converge:
        \[
            \eng^{av}\left(\x^{(0)}, \ldots, \x^{(K)}\right) = (1 - O(1/K)) \cdot \eng(\b^*),
        \]
        \item Diversity approaches $\d$ exponentially fast:
        \[
            \divers(\x^{(k)}) \ge \d - \frac{1}{\exp(\Omega(k))},
        \]
        \item The policy $\b^*$ achieves approximately-optimal engagement. That is, if there is a strategy $\b^{(0)}, \ldots, \b^{(K)}$ inducing a sequence $\y^{(0)}, \ldots, \y^{(K)}$ such that $\divers(\y^{(k)}) \ge \d$ for all $k$, then
        \[
            \eng^{av}\left(\y^{(0)}, \ldots, \y^{(K)}\right) \le (1 + O(1/K)) \cdot \eng^{av}\left(\x^{(0)}, \ldots, \x^{(K)}\right).
        \]
    \end{enumerate}
\end{theorem}
The proof of \Cref{thm:main} relies on the following simple lemma, a consequence of Gelfrand's formula~\citep{rudin1991functional},
the main argument of which was proved by an anonymous user on Stackexchange,\footnote{\texttt{https://math.stackexchange.com/questions/2561701/bound-on-the-norm-of-a-matrix-power}} although variations are clearly known in the literature. We nonetheless include the entire argument for completeness.
\begin{lemma}\label{lem:matrix}
    There are constants $\lambda > 0$ and $\gamma \in (0, 1)$ depending only on $G$ and $\p$ such that for each type matrix $\A_t$, any injection policy $\b_t$, and any power $\ell \ge 0$, $\sum_{\ell = k}^\infty \|\A_t^\ell \b_t\|_1 \le \lambda \gamma^k$.
\end{lemma}
\begin{proof}
Gelfand's formula implies that for any type $t$, $\lim_{\ell \to \infty} ||\A_t^\ell||_1^{\frac{1}{\ell}} = \rho(\A_t)$, where $\rho(\A_t)$ is the spectral radius of $\A_t$ (see, e.g., \citet{rudin1991functional}, Theorem 10.35). As $\rho(\A_t) < 1$ for all types $t$, we can choose $\gamma \in (\max_t{\rho(\A_t)},1)$ and doing so will imply $\lim_{\ell \to \infty} ||\A_t^\ell||_1^{\frac{1}{\ell}} < \gamma$. means that for each type $t$, for sufficiently large $\ell$, it holds that $\|\A_t^\ell\|_1 < \gamma^\ell$. Hence, we can choose $M > 0$ large enough so that $||\A_t^\ell ||_1 < M \gamma^\ell$ for all $\ell$ and $t$. Finally, observing that $\|\b_t\|_1 < n$ since it is a valid injection strategy, we have,
\begin{align*}
    \sum_{\ell = k}^\infty ||\A_t^\ell \b_t||_1
    &\leq  \sum_{\ell = k}^\infty ||\A_t^\ell||_1 \cdot || \b_t||_1\\
    &< n \sum_{\ell = k}^\infty M \gamma^{\ell}\\
    &= \left(M n (1-\gamma)^{-1} \right) \gamma^k.
\end{align*}
Choosing $\lambda = \left(M n (1-\gamma)^{-1} \right)$ completes the proof.
\end{proof}
\begin{proof}[Proof of \Cref{thm:app-optimal}]
    Fix an instance $G = ([n], E)$, retweet probabilities $\p$, and a value $\d \le 1/T$. Fix a solution $\b^*$ to the $\d$-diversity program, a time horizon $K$, and induced states $\x^{(0)}, \ldots, \x^{(K)}$. Fix the corresponding constants $\lambda$ and $\gamma$ from \Cref{lem:matrix}. Recall that $\x_t^{(k)} = \sum_{\ell = 0}^k \A_t^\ell \b^*$.
    
    We first consider part (1). Recall that $\x(\b^*) = \sum_{k = 0}^\infty \A^\ell$ and hence $\x(\b^*)$ dominates $\x^{(k)}$ component-wise. Since $\eng(\x)$ is monotonic in the components of $\x$, this implies that $\eng(\b^*) \ge \eng(\x^{(k)})$ for all $k$, so
    \[
    \frac{1}{K + 1} \sum_{k = 0}^K \eng(\x^{(k)}) \le \frac{1}{K + 1} \sum_{k = 0}^K \eng(\b^*) =  \eng(\b^*).
    \]
    In the degenerate case where $\eng(\b^*) = 0$, (1) immediately follows as both sides are equal to $0$. Hence, we now consider the case where $\eng(\b^*) > 0$. Notice that $\eng(\b^*)$ does not depend on $K$ and is hence a constant in the $O()$ formula, so, rearranging the statement, it suffices to show that 
    \[
        \eng(\b^*) - \eng^{av}\left(\x^{(0)}, \ldots, \x^{(K)}\right) = O(1/K).
    \]
    For the rest of the proof, it will be useful to observe that $\x^{(k)}$ converges to $\x(\b^*)$. More formally, using \Cref{lem:matrix}, we have that for all types $t$ and times $k$,
    \begin{equation}\label{ineq:converge}
        \norm*{\x(\b^*)_t - \x_t^{(k)}}_1 = \norm*{\sum_{\ell = 0}^\infty \A^\ell_t \b^*_t - \sum_{\ell = 0}^k \A^\ell_t \b^*_t}_1 =  \norm*{\sum_{\ell = k + 1}^\infty \A^\ell_t \b^*_t}_1 \le \lambda\gamma^{k + 1}
    \end{equation}
    Additionally, we observe that $\inner{\p_t, \x_t} \le \|\x_t\|_1$ for all $\x$ because each component of $\p_t < 1$. By combining these facts and expanding definitions, it follows that:
    \begin{align*}
        \eng(\b^*) - \eng^{av}\left(\x^{(0)}, \ldots, \x^{(K)}\right)
        &= \sum_t \inner{\p_t, \x(\b^*)_t} - \frac{1}{K + 1}\sum_{k = 0}^K \sum_t \inner{\p_t, \x^{(k)}_t}\\
        &= \frac{1}{K + 1}\sum_{k = 0}^K \sum_t \inner{\p_t, \x(\b^*)_t}- \frac{1}{K + 1}\sum_{k = 0}^K \sum_t \inner{\p_t, \x^{(k)}_t}\\
        &= \frac{1}{K + 1}\sum_{k = 0}^K \sum_t \inner{\p_t, \x(\b^*)_t - \x_t^{(k)}}\\
        &\le \frac{1}{K + 1}\sum_{k = 0}^K \sum_t \norm*{\x(\b^*)_t - \x_t^{(k)}}_1\\
        &\le \frac{1}{K + 1}\sum_{k = 0}^K \sum_t \lambda\gamma^{k + 1}\\
        &= \frac{t}{K + 1}\sum_{k = 0}^K  \lambda\gamma^{k + 1}\\
        &\le\frac{t}{K + 1}\sum_{k = 0}^\infty  \lambda\gamma^{k + 1}\\
        &= \frac{t \lambda \gamma}{(1 - \gamma)(K + 1)} = O(1/K).
    \end{align*}
    
    Next, we consider part (2), which follows more straightforwardly from \Cref{ineq:converge}. Since $\divers(\x(b^*)) \ge \d$ by assumption, each component of $\x(b^*) \ge t$. Since no component can differ by more than the $L_1$ distance between vectors, each component of $\x_t^{(k)}$ is at least $\d - \lambda\gamma^{k + 1}$ by Inequality~\eqref{ineq:converge}. Since $\gamma < 1$, $\lambda\gamma^{k + 1} = \frac{1}{\exp(\Omega(k))}$, as needed.
    
    We now move on to part (3). Fix a $\d$-diverse strategy $\b^{(0)}, \ldots, \b^{(K)}$. Consider $\b^{av} = \frac{1}{K + 1}\sum_{k = 0}^K \b^{(k)}$, the average of the time-dependent injections. First, observe that $\b^{av}$ is, in fact, a valid injection policy (i.e., nonnegative with no user shown more than one unit) since it is the linear combination of valid injection policies. We use $\b^{av}$ to more directly compare the time-dependent strategy to $\b^*$.

    We begin by showing that $\x(\b^{ab})$ component-wise dominates $\frac{1}{K + 1} \sum_{k = 0}^{K} \y^{(k)}$, i.e., $x(\b^{ab})_{ti} \ge \frac{1}{K + 1} \sum_{k = 0}^{K} y^{(k)}_{ti}$ for all $t$ and $i$. To that end, we unravel the recursive definition of $\y^{(k)}_t$. We have
    \[
        \y^{(k)}_t = \sum_{\ell = 0}^k (\A_t)^{k - \ell} \b^{(\ell)}_t.
    \]
    Plugging that into the linear combination,
    \[
        \frac{1}{K + 1} \sum_{k = 0}^{K} \y^{(k)}_t = \frac{1}{K + 1} \sum_{k = 0}^{K} \sum_{\ell = 0}^{k} (\A_t)^{k - \ell} \b^{(\ell)}_t.
    \]
    Notice that a term $\A_t^{a}\b_t^{(b)}$ with a specific combination of $a$ and $b$ can only appear at most once, only when the outside sum has $k = a + b$. Hence, since all the summands are nonnegative, we have 
    \begin{align*}
         \frac{1}{K + 1} \sum_{k = 0}^{K} \y^{(k)}_t
        &\le \frac{1}{K + 1} \sum_{\ell = 0}^{K} \sum_{k = 0}^{K} (\A_t)^{\ell} \b^{(k)}_t\\
        &=   \sum_{\ell = 0}^{K} (\A_t)^{\ell}\left(\frac{1}{K + 1}\sum_{k = 0}^{K}  \b^{(k)}_t\right)\\
        &=\sum_{\ell = 0}^{K} (\A_t)^{\ell} \b^{av}_t\\
        &\le \sum_{\ell = 0}^{\infty} (\A_t)^{\ell} \b^{av}_t\\
        &= \x(\b^{av})_t
    \end{align*}
    with $\le$ defined component-wise. Using this, we have that
    \[
        \eng(\b^{av}) \ge \frac{1}{K + 1}\sum_{k = 0}^K \eng(\y^{(k)}) = \eng^{av}\left(\y^{(0)}, \ldots, \y^{(K)}\right).
    \]
    Further, notice that since for each $k$, $\divers(\y^{(k)}) \ge \d$, $\divers\left(\frac{1}{K + 1}\sum_{k = 0}^{K} \y^{(k)}\right) \ge \d$, so again by the component-wise domination, $\divers(\x(b^{av})) \ge \d$. This implies that $\b^{av}$ is a feasible solution to the $\d$-diversity program. Hence, by the optimality of $\b^*$, $\eng(\b^{av}) \le \eng(\b^*)$. Putting it all together, we have
    \[
        \eng^{av}\left(\y^{(0)}, \ldots, \y^{(K)}\right) \le \eng(\b^{av}) \le \eng(\b^*) \le (1 + O(1/K))\eng^{av}\left(\x^{(0)}, \ldots, \x^{(K)}\right),
    \]
    where the last inequality follows from part (1).
\end{proof}

\section{Theoretical Bounds on the Engagement-Diversity Tradeoff}\label{sec:theory}
We now turn to providing bounds on the cost of $\d$-diversity. In order to prove upper bounds, rather than focusing on optimal injection policies, we consider algorithms that, while not optimal, are easier to analyze. We begin this section by defining two. To do so, recall that $c_{ti}$, the coefficient in the optimal programs, represents engagement generated in the limiting state by injecting a unit of type $t$ tweet to user $i$. For each user $i$, let $f_i \in \argmax_t c_{ti}$ be a tweet type generating maximal engagement. Additionally, recall that $\engopt = \sum_i \max_t c_{t, i}$, achieved by injecting a unit of $f_i$ to each user $i$.

\begin{definition} The \emph{$\d$-uniform policy} for each user $i$ injects $\d$ of each type $t$ and spends the remaining $1 - T\d$ budget on $f_i$. More formally, $b_{ti} =  \d$ for $t \ne f_i$ and $b_{f_i i} = 1 - (T - 1)\d$. 
\end{definition}
This policy is $\d$-diverse as it directly injects at least $\d$ of every type to all users. Using the $\d$-uniform policy, we can immediately derive a worst-case bound on $\dcost(G, \p)$ for all graphs $G$ and retweet probabilities $\p$. Indeed, regardless of the underlying values, by injecting $1 - (T - 1)\d$ units of $f_i$, we have that it achieves engagement at least 
$$
(1 - (T - 1)\d) \sum_i \max_t c_{ti} = (1 - (T - 1)\d)\engopt.
$$
Since $\divopt$ must be at least the engagement of this policy, we have 
\begin{equation}
\label{eq:worst}
\dcost(G, \p) = 1 - \frac{\divopt}{\engopt} \le (T - 1)\d
\end{equation}
as a worst-case bound. 

Note that the bound of \Cref{eq:worst} is, in some cases, tight. Indeed, consider an empty graph where all users have a positive retweet probability for only one type. In an empty graph, the limiting state is exactly equal to the injection policy. Hence, to achieve $\d$-diversity, it is necessary to inject $\d$ of all types to everybody, but this means only a $(1 - (T - 1)\d)$ of the policy can be spent on types from which users derive any engagement.

However, to get beyond this worst-case bound, we need slightly more intricate policies to analyze. The next policy is based on the following idea: Suppose we wish to inject tweets such that, in the limiting state, every user sees \emph{exactly} $\d$ of each type. Notice that computing this policy is not, in fact, too difficult. For a fixed type $t$, the injected $\b_t^{ex}$ would need to satisfy
\[
    \A^*_t \b^{ex}_t = \delta \mathbf{1}.
\]
Expanding the definition $\A^*_t = (\I - \A_t)^{-1}$ and multiplying on both sides, we have that
\[
    \b^{ex}_t = \delta(\mathbf{1} - \A_t \mathbf{1}).
\]
Hence, we have that
\[
    b^{ex}_{ti} = \delta \left( 1 - \sum_j A_{tij}\right).
\]
It is not immediately obvious that the collection $\b^{ex}$ is a valid injection policy; however, we can prove this is the case. Indeed, notice each $\b^{ex}_{ti} \le \delta$ because entries in $\A$ are nonnegative, so by our restriction that $\delta \le 1/T$, $\sum_i \b^{ex}_{ti} \le 1$. Further, our normalization by $\fol(i)$ in $\A_t$ ensures that $\sum_j \A_{tij} \le \max p_{ti} \le 1$, so each $\b^{ex}_{ti} \ge 0$.

Since this sum $\sum_j \A_{tij}$ will come up frequently, we define $\inc_{ti} = \sum_j \A_{tij}$. Intuitively, $\inc_{ti}$ is the ``incoming weight'' from all the users $i$ follows. If all users were shown exactly 1 unit of type $t$ at time $0$, then, in the next time step, $i$ would see $\inc_{ti}$ units. It follows that, in a steady state where all users see $\delta$ of type $t$, to ensure they see $\delta$ of type $t$ at the next time step, they must be injected $\delta(1 - \inc_{ti})$. 

We can now use $\b^{ex}$ to define a new injection policy.

\begin{definition} The \emph{$\d$-exact policy} first injects $\b^{ex} = \d(1 -\inc_{ti})$ of type $t$ to each user $i$ and then spends the remaining $1 - \d\sum_t(1 - \inc_{ti})$ on $f_i$. More formally, $b_{ti} = \d(1 - \inc_{ti})$ for $t \ne f_i$ and $b_{f_i i} = 1 - \d(T - 1 - \sum_{t \ne f_i} \inc_{ti})$.
\end{definition}

Since the $\d$-exact policy is injecting only more than $\b^{ex}$, clearly all users see at least $\delta$ of each type in the limit, so it must be $\d$-diverse. Using the $\d$-exact policy, we derive what we view as our main theoretical result.
\begin{theorem}\label{thm:main}
    Fix constants $\alpha \le \beta$ with $\beta > 0$. For all graphs $G$ and retweet probabilities $\p$ such that for each user $i$
    \begin{enumerate}
        \item their average retweet probability is at least $\alpha$, i.e., $\frac{1}{T} \sum_{t} p_{ti} \ge \alpha$, and
        \item their maximum retweet probability is at most $\beta$, i.e., $\max_{t} p_{ti} \le \beta$,
    \end{enumerate}
    it holds that
    \[
        \dcost(G, \p) \le \min\left\{T\d\left(1 - \frac{\alpha}{\beta}\right), (T - 1)\delta\right\}.
    \]
    Further, for all $\alpha\leq \beta$ with $\beta>0$, there are instances $G$ and $\p$ satisfying (1) and (2) such that the inequality is tight.
\end{theorem}
One special case of particular interest is when users are homogeneous, that is, all having the same retweet probabilities, say $p_1 \ge \cdots \ge p_T$. In this case, as long as we were not in the degenerate case where $p_t = 0$ for all $t$, the bound simplifies to:
\[
    \dcost(G, \p) \le \delta\left( \sum_{t \ne 1}\left( 1 - \frac{p_t}{p_1}\right) \right).
\]
This bound is at least as strong as the theoretical worst case and strictly stronger when it is not the case that $p_2 = \cdots = p_t = 0$.

We also note that, as the proof of \Cref{thm:main} shows, the bound is tight in an empty graph (with no edges). It is also not hard to establish the upper bound in such a graph. The power of \Cref{thm:main}, then, lies in showing that the engagement-diversity tradeoff is no worse when generalizing an empty graph to an arbitrary social network with elaborate retweet dynamics. 


\begin{proof}[Proof of \Cref{thm:main}]
    Fix some $\alpha$ and $\beta$, a graph $G$, and retweet probabilities $\p$ satisfying the theorem requirements. Notice that the upper bound of $(T - 1)\delta$ follows from the theoretical worst-case discussed above. So for this proof, we show an upper bound of $T\delta(1 - \frac{\alpha}{\beta})$.
    To prove the theorem, we analyze the engagement derived from the $\d$-exact policy. We partition the users $V = I \sqcup O$ where $I$ is the set of \emph{inside} users that follow at least one other and $O$ is the set of \emph{outside} users that do not follow anybody. For each outside user $i \in O$, since they do not follow anybody, the policy spends the entire $\d$ on each type in the first part, and then spends the remaining $1 - T\d$ on $f_i$. For the inside users $i \in I$, as the average retweet probability for each user they follow is at least $\alpha$, the sum of all incoming edges (even after scaling down by the number of followers) is at least $T\alpha$. Hence, there is a remaining budget of at least $1 - T\d  + T\d \alpha$ to spend on $f_i$.

    Next, let us consider the engagement derived. Recall that $\engopt = \sum_{i \in [n]} \max_t c_{ti}$. We define $E_I = \sum_{i \in I} \max_t c_{ti}$ and $E_O = \sum_{i \in O} \max_t c_{ti}$ to be engagements derived from injecting to inside and outside users respectively. Note that $\engopt = E_I + E_O$. Now let us consider the engagement of the $\d$-exact policy. Notice that by the first part alone, each user sees $\d$ of each type, and by the $\alpha$ lower bound, they must derive $\d T \alpha$ engagement from this. In the second part, outside users $i$ contribute $(1 - T\d) \max_t c_{ti}$ and inside users $i$ contribute at least $(1 - T\d + T \d \alpha) \max_t c_{ti}$. Putting this together, we have that the total engagement is at least
    \begin{equation}\label{eq:strat-engagement}
        \underbrace{nT\d \alpha}_{\text{First part}} + \underbrace{(1 - T\d)E_O}_{\text{Outside user second part}} + \underbrace{(1 - T\d + T\d\alpha) E_I}_{\text{Inside user second part}}
    \end{equation}
    Our goal then is to show that \eqref{eq:strat-engagement} is at least
\[
    \left(1 - T\d\left(1 - \frac{\alpha}{\beta}\right) \right) \engopt.
\]
We begin by showing that
\begin{equation}\label{ineq:beta}
    (1 - \beta) E_I + E_O \le \beta n.
\end{equation}
To that end, we consider a modified instance $(G', \p')$ where the graph remains the same, so $G = G'$, but we set $\p'$ such that $p'_{ti} = \beta$ for all $t$ and $i$. Notice that this has only increased the values of $\p$. Let $\c'$ be the corresponding $\c$ values and define $E'_I$ and $E'_O$ analogously using $\c'$. Since increasing retweet probabilities can only increase the values $c_{ti}$, this also holds for $E_I$ and $E_O$, so $(1 - \beta)E_I + E_O \le (1 - \beta)E'_I + E'_O$. Hence, it is sufficient to show that $(1 - \beta)E'_I + E'_O \le \beta n$.

In this modified instance, all types are symmetric, so $c'_{t_1i} = c'_{t_2i}$ for all types $t_1$ and $t_2$. Hence, the vector $(\max_t c_{t1}, \ldots, \max_t c_{tn}) = \c'_1$, i.e., the vector for type $1$ (or any $\c_t$ vector for that matter). Let $\A'_1$ be the type matrix of type $1$ in the modified instance, so $A'_{1ij} = \frac{\beta}{\fol(i)}$ if $i$ follows $j$ and $0$ otherwise. Recall that $(\c'_1)^{\top} = (\beta\mathbf{1})^{\top} (\I - \A_1)^{-1}$.

Let $\b'_1$ be the vector where $\b'_{1i} = 1$ if $i$ is an outside user and $\b'_{1i} = 1 - \beta$ if $i$ is an inside user. The value $(1 - \beta)E'_I + E'_O$ is exactly equal to
\[
    \inner{\c'_1, \b'_1} = (\c'_1)^{\top}\b'_1 = (\beta\mathbf{1})^{\top} (\I - \A_1)^{-1}\b'_1.
\]
Another interpretation of this is the engagement derived in the limiting state after injecting $(1 - \beta)$ of type $1$ to inside users and $1$ unit of type $1$ to outside users. 

We first claim that $(I - \A'_1)^{-1} \b'_1 = \mathbf{1}$, the all ones vector. Indeed, such a solution is the unique $\y$ that satisfies $(I - \A'_1)\y = \b'_1$. Notice that for the all ones vectors, for outside users, the $i$'th row of $\A'_1$ is all $0$s, so the term is $0$, and for inside users, the $i$'th row has $\fol(i)$ number of terms each with value $\frac{\beta}{\fol(i)}$, so the corresponding component is $1 - \beta$. Therefore, $\mathbf{1}$ is the limiting state, and the engagement is exactly $(\beta\mathbf{1})^{\top}\mathbf{1} = n\beta$. This equality implies $(1 - \beta)E'_I + E'_O \le \beta n$, as needed.

From Inequality~\eqref{ineq:beta}, by simple algebra, we get \[n + E_I \ge \frac{E_I + E_O}{\beta} = \frac{\engopt}{\beta}.\]

Using this inequality, we have
\begin{align*}
\phantom{{}={}}  nT\d\alpha + (1 - T\d)E_O + (1 - T\d + T\d\alpha) E_I
&=nT\d\alpha + T\d\alpha E_I + (1 - T\d)\engopt\\
&= T\d\alpha (n + E_I) + (1 - T\d)\engopt\\
&\ge \frac{T\d \alpha \engopt}{\beta}  + (1 - T\d)\engopt\\
&= \left(1 - T\d + \frac{T\d \alpha }{\beta}\right) \engopt\\
&=\left(1 - T\d \left(  1 - \frac{\alpha}{\beta} \right)\right)\engopt,
\end{align*}
as needed.

To show tightness, consider an empty graph $G$ where, for all users $i$,
\[p_{ti} = 
\begin{cases}
    \beta & \text{ if } t = 1\\
    \max\left\{\alpha - \frac{\beta - \alpha}{T - 1}, 0\right\} & \text{ if } t \ne 1.
\end{cases}
\]
For ease of notation, we let $\gamma = \max\left\{\alpha - \frac{\beta - \alpha}{T - 1}, 0\right\}$.
Notice that the maximum retweet probability is $\beta$ and
 $\alpha - \frac{\beta - \alpha}{T - 1}$ is the exact value for other types to get the average retweet probability to $\alpha$, \[ 
\frac{\beta + (T - 1)\left( \alpha - \frac{\beta - \alpha}{T - 1}\right)}{T} = \frac{\beta + (T - 1)\alpha - \beta + \alpha}{T} = \alpha,
\]
so the true average is  
$$
\frac{\beta + (T - 1)\gamma}{T} = \max\left\{\frac{\beta}{T}, \alpha\right\} \ge \alpha.
$$
Further, notice that the since the graph is disconnected, the limiting state is exactly the injection policy. In this case, one policy is to inject only type $1$ which gives engagement $n\beta$. Hence, $\engopt \ge n\beta$. To be $\delta$-diverse, a policy must inject at least $\delta$ of each type to all users and can therefore inject at most $1 - (T - 1)\delta$ to type $1$. Hence,
\[
    \divopt \le \left(\underbrace{\beta\left(1 - (T - 1)\delta\right)}_{\text{type $1$}} +\underbrace{(T - 1)\delta \gamma}_{\text{other types}} \right)n.
\]
Plugging these inequalities in, we have
\begin{align*} 
    \dcost(G, \p)
    &= 1 - \frac{\divopt}{\engopt}\\
    &\ge 1 - \frac{\left(\beta\left(1 - (T - 1)\delta\right) +(T - 1)\delta \gamma \right)n.}{\beta n}\\
    &= 1 - \left(1 - (T - 1)\delta + \frac{(T - 1)\delta\gamma}{\beta} \right)\\
    &= (T - 1)\delta - \frac{(T - 1)\delta\gamma}{\beta}\\
    &= (T - 1)\delta \left(1 - \frac{\gamma}{\beta}\right)\\
    &= T \delta \frac{T - 1}{T} \left(1 - \frac{\gamma}{\beta}\right)\\
    &= T \delta \left(1 -  \frac{1}{T} - \frac{(T - 1)\gamma}{T \beta}\right)\\
    &= T \delta \left(1 -  \frac{\beta + (T - 1)\gamma}{T} \cdot \frac{1}{\beta}\right)\\
    &= T \delta \left(1 -  \max\left\{\alpha, \frac{\beta}{T} \right\} \frac{1}{\beta}\right)\\
    &= T \delta \left(1 -  \max\left\{\frac{\alpha}{\beta}, \frac{1}{T} \right\} \right)\\
    &= T \delta \min\left\{1 - \frac{\alpha}{\beta}, \frac{T - 1}{T} \right\}\\
    &= \min\left\{T\d\left(1 - \frac{\alpha}{\beta}\right), (T - 1)\delta\right\},
\end{align*}
as needed.
\end{proof}




\section{The Engagement-Diversity Tradeoff in Practice}\label{sec:experiments}

This section aims to use data to infer practical settings of $G$ and $\p$ and gain an empirical understanding of the cost of $\delta$-diversity.

\subsection{Data Processing}

We reconstruct the social network ($G$) of users and their retweet probabilities ($\p$) with respect to different tweet types  from a Twitter dataset, which has been analyzed and evaluated in previous studies~\citep{garimella1,garimella2,garimella3}.
The dataset consists of tweets posted during one week around specific events that spurred an increased interest in different political issues. Here, we specifically consider users that tweeted in the context of abortion around June 30, 2016, when the U.S. Supreme court struck down Texas abortion restrictions.
We refer the curious reader to \citet{garimella1} for in-depth details about the dataset.

\begin{figure}[t]
    \centering
    \includegraphics[width=.45\textwidth]{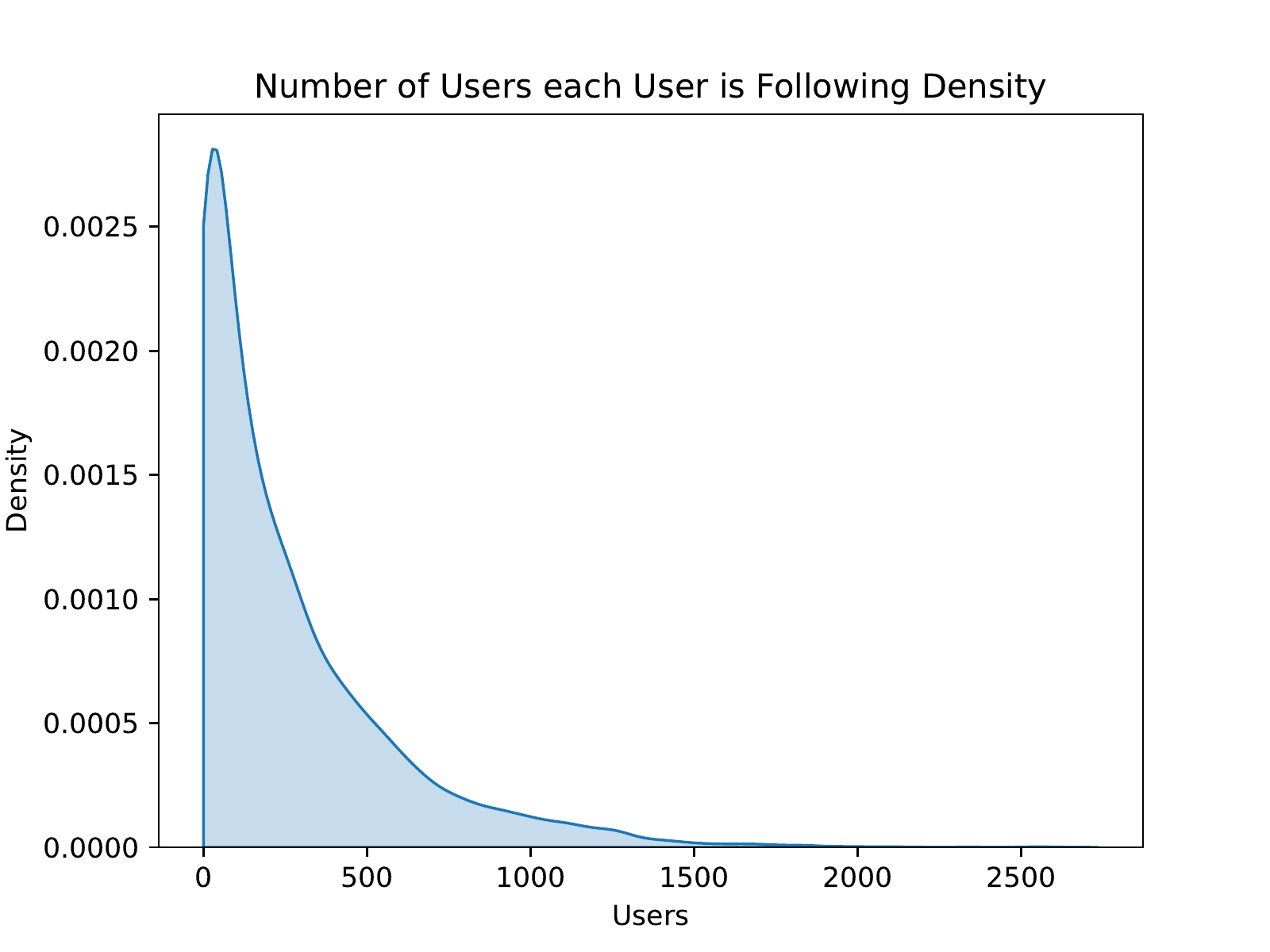}
    \includegraphics[width=.45\textwidth]{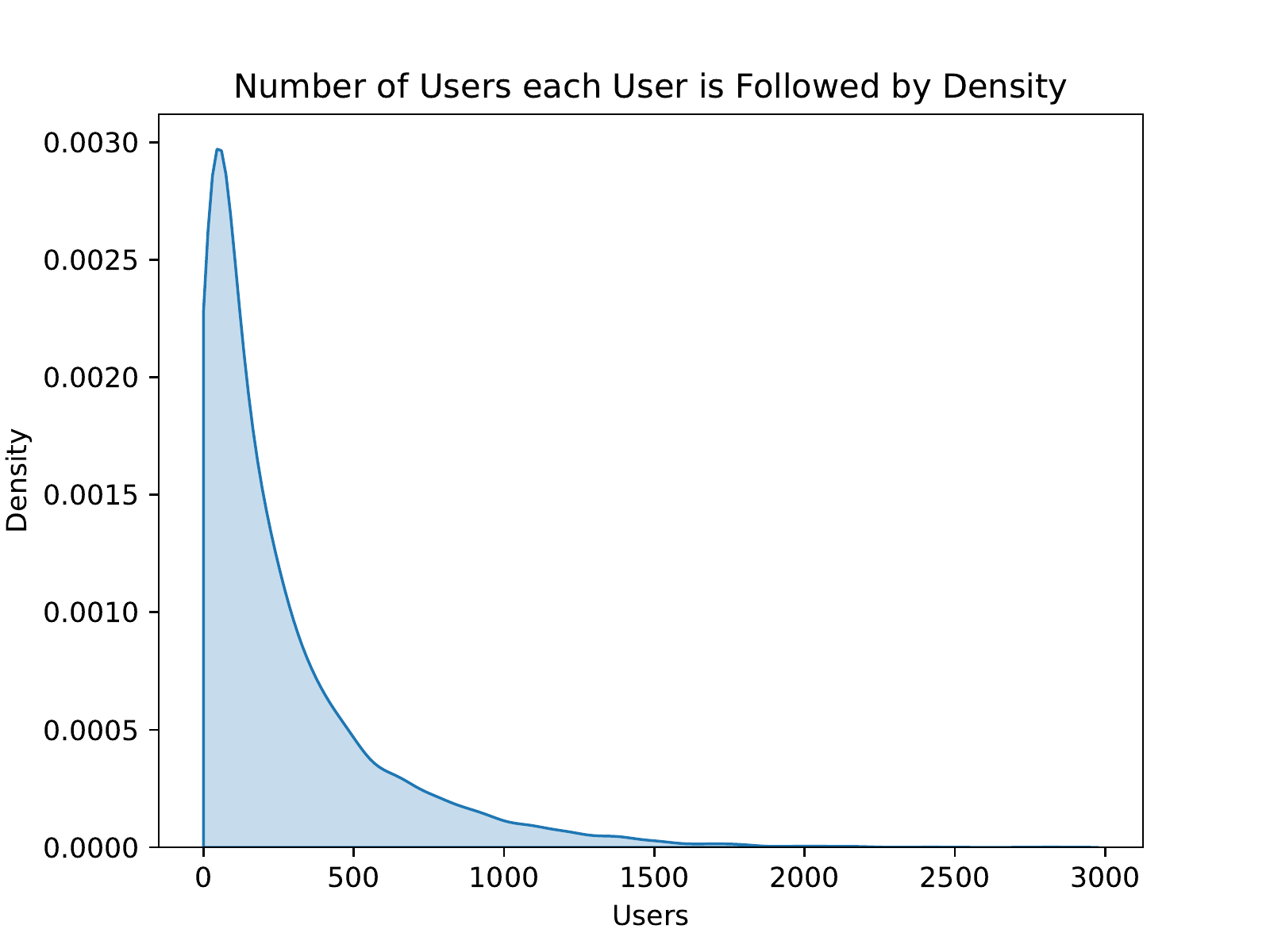}
    \caption{Densities of the number of users each user follows and is followed by in the dataset. Densities were computed using kernel density estimation.}
    \label{fig:degrees}
\end{figure}

Briefly, the dataset contains tweets along with information including the tweet text and metadata such as the ID of the Twitter user that tweeted (or retweeted) and hashtags mentioned in the tweet.
Additionally, each user's social relations on Twitter were recorded, i.e., which other users they follow. This allows us to directly reconstruct the follower graph $G$. In all, there are 7,284 users and a total of 1,880,679 edges, so each user is, on average, following roughly 258 others. The distributions of how many users each user follows and is followed by can be found in \Cref{fig:degrees}.

We use hashtags to classify the tweets in the dataset into types, allowing us to later infer retweet probabilities. In particular, as the number of distinct hashtags is extremely large (more than 695,000), we restrict the analysis to the 2,000 most common hashtags. Using this, we construct a hashtag network, where a link between two hashtags indicates that these hashtags appeared in the same tweet and are therefore related~\cite{amati2021topic,pervin2015hashtag}.
We then use this network to cluster hashtags into a more manageable number of ``types.'' Specifically, we use the Louvain algorithm~\cite{blondel2008fast} on the hashtag network to extract its community structure, where each community of hashtags defines a specific type. In all, this gives us a classification of hashtags into four distinct types.

We then use hashtag occurrences as a proxy for the number of tweets of a certain type.
We first compute the number of times a user retweets a type by counting the times a corresponding hashtag appeared. Note that, generally, a single tweet may be considered part of multiple types or increase the count of a particular topic by more than one, as the tweet may contain multiple hashtags. To determine the number of tweets of a particular type \emph{seen} by a user, we count the number of corresponding hashtag occurrences in their neighbors' tweets, both original and retweets.

Finally, we use two methods to infer retweet probabilities using the counts. First, we simply take the proportion of retweets divided by the number of tweets seen of a type. We call these ``mode probabilities.'' For the second, we assume that the dataset is just a single observation of a user's retweet probability and use Bayesian updating. Specifically, we assume an independent $\text{Beta}(1, 100)$ prior on users' retweet probabilities which reasonably closely corresponds to the distribution of retweet probabilities observed. We then do Bayesian updating to get a posterior distribution on retweet probabilities; this has the convenient property that if a user retweets $r$ out of a total of $s$ seen, the posterior is $\text{Beta}(1 + r, 100 + s)$. 

We take two samples from the Beta distribution for each user. Together with the mode probabilities, we end up with three instances that share the same graph but have distinct retweet probabilities.

\subsection{Experiments}

For each of the three instances, analyze the engagement-diversity tradeoff as the initial retweet probabilities are scaled up. To do this, we directly multiply all inferred $p_{it}$ probabilities by a constant. This is done for factors $1$, $3$, $10$, and $30$ ($1$, of course, being the original inferred probabilities). After scaling, we always cap the maximum retweet probability to $.99$. The capping ensures both that the probabilities are consistent with their modeling definition and that the linear system converges.

For each instance, we compute the necessary values to run the engagement-optimal and $\delta$-diverse linear programs, i.e., compute the type matrices and find their limiting value. From this, we can immediately solve the first LP to obtain $\engopt$. Next, we actually solve the LPs for $10$ evenly spaced values of $\delta$, $\frac{1}{10T}, \frac{2}{10T}, \ldots, \frac{1}{T}$ which gives us $\divopt$ for these values. In all our experiments, the number of types is $T = 4$. With these values, we can plot $\dcost(G, \p) = 1 - \frac{\divopt}{\engopt}$. In the plots, we also include the theoretical worst-case bound of $(T - 1)\delta$.

\begin{figure}[t]
    \centering
    \includegraphics[width=.45\textwidth, valign=m]{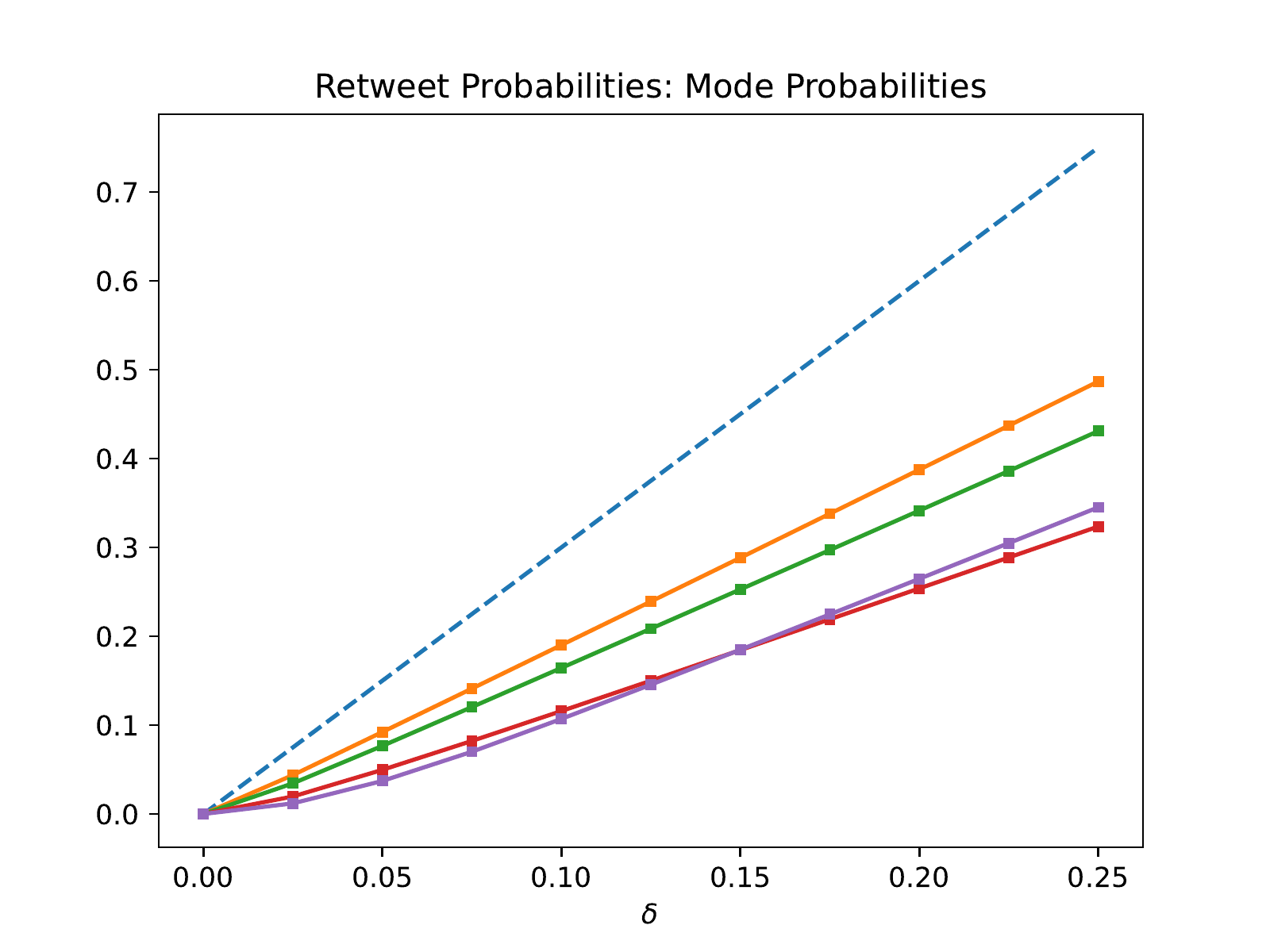} \includegraphics[width=.45\textwidth, valign=m]{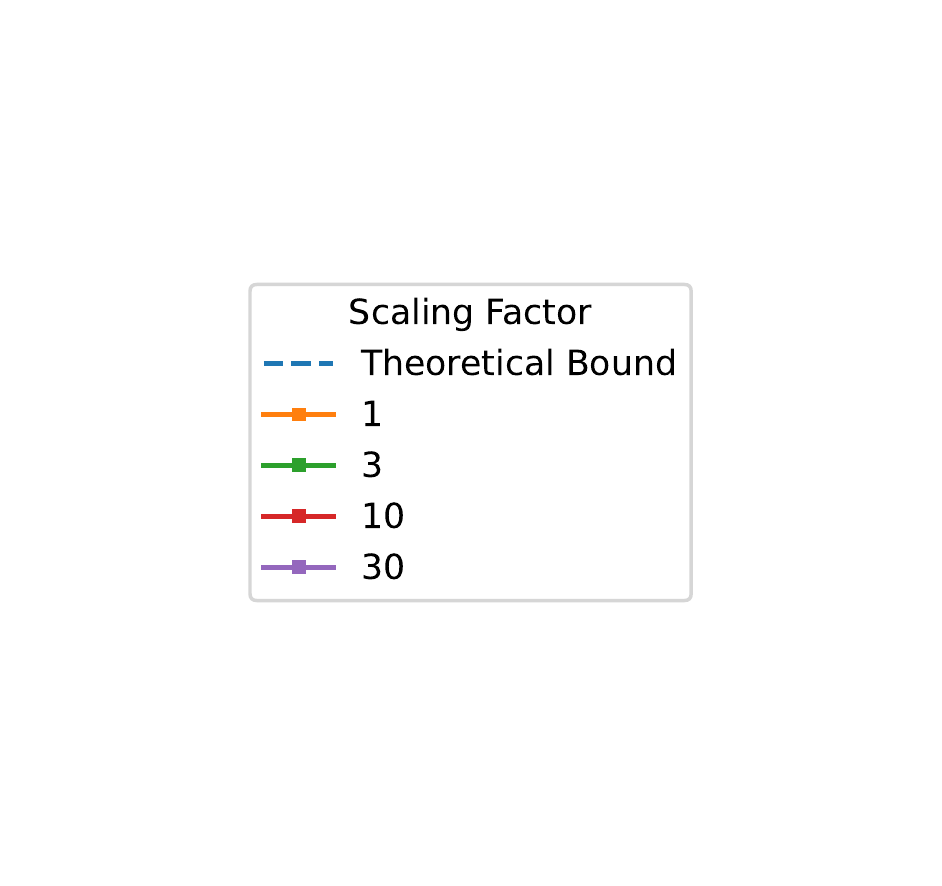}\\
    \includegraphics[width=.45\textwidth, valign=m]{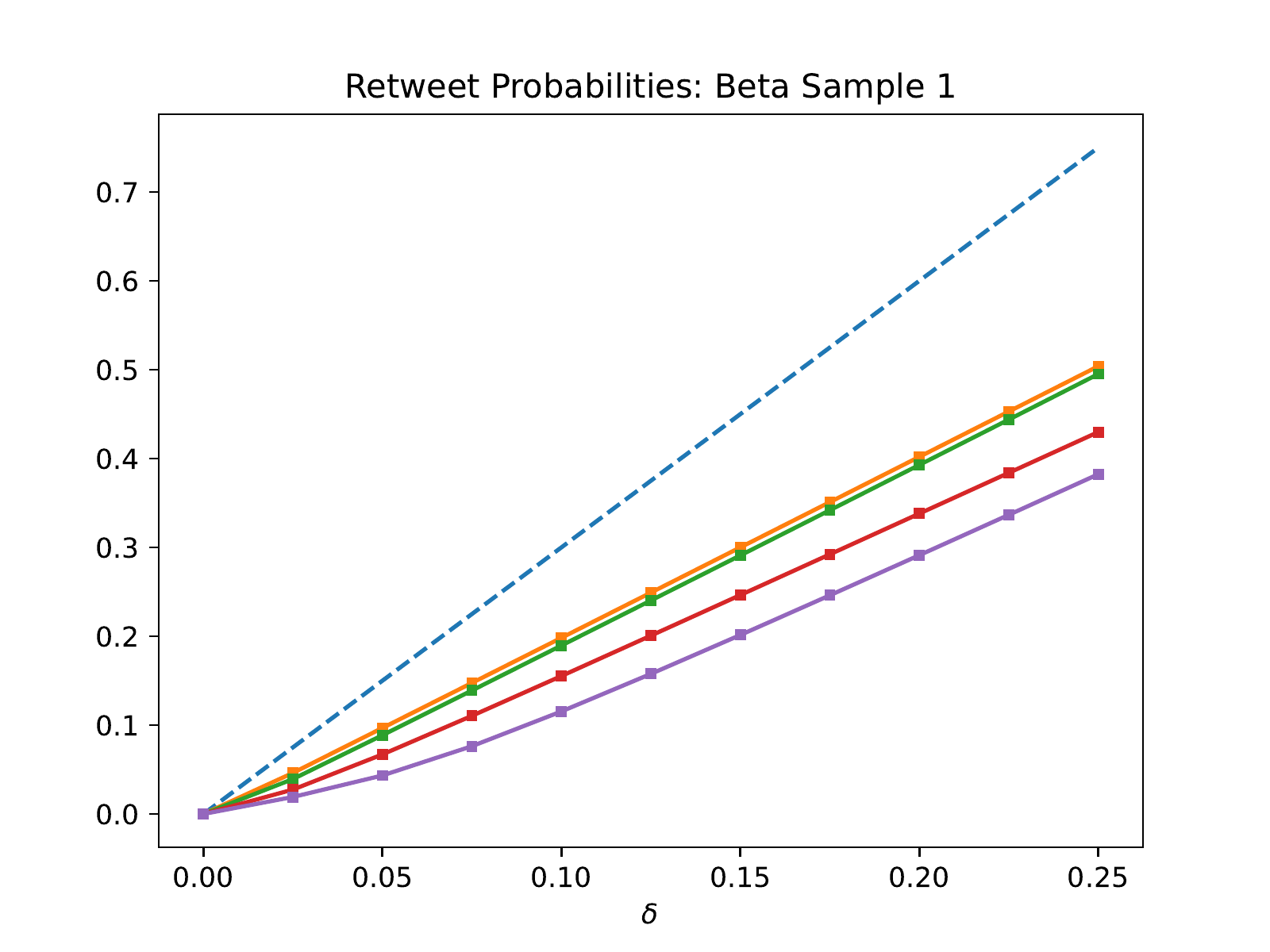}
    \includegraphics[width=.45\textwidth, valign=m]{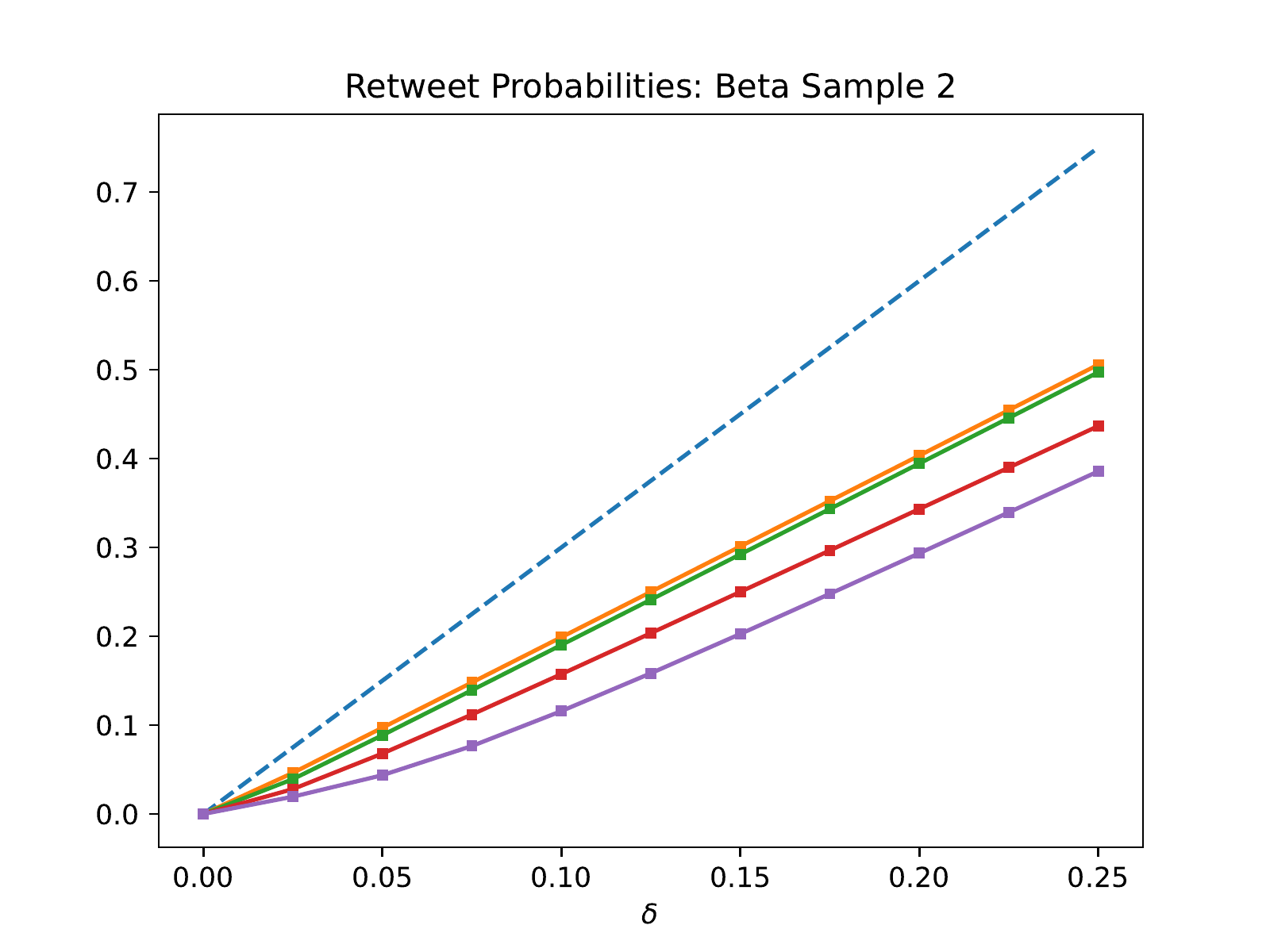}
    \caption{Plots of $1 - \frac{\divopt}{\engopt}$ in various settings. Mode probabilities were inferred by directly calculating the ratio between retweeted and inferred. Beta samples were samples from a posterior distribution of retweet probabilities. The blue line in each diagram indicates the theoretical lower bound from \Cref{sec:theory}; it is equal to $(T - 1) \delta$ where $T = 4$ in all settings. Colored lines indicate different scaling factors of the original retweet probabilities. Values of $1 - \frac{\divopt}{\engopt}$ are computed for values of $\frac{i}{10 \cdot T}$ for $i = 0, \ldots, 10$. }
    \label{fig:plots}
\end{figure}

Due to the large size of the dataset, these experiments are computationally intensive. All LPS were solved using Gurobi on an Amazon Web Services (AWS) instance with 128 vCPUs of a 3rd Gen AMD EPYC running at 3.6GHz equipped with 1TB of RAM. Giving a Gurobi solver three threads, it takes on the order of 30 hours to compute the optimal values of the $10$ LPs with $\delta = \frac{1}{10T}, \ldots, \frac{1}{T}$. We ran all of our experiments in parallel, which used approximately 500 GB of RAM during the solve.

\subsection{Results}

Results are shown in \Cref{fig:plots}. The different colored lines correspond to the different scaling factors along with the theoretical lower bound derived in \Cref{sec:theory} of $(T - 1)\delta$.

As a general way to interpret the results, we consider the perspective of a policy-maker. If they were willing to sacrifice 5\% of engagement in order to boost diversity, with small probabilities, they may expect to be able to get $0.03$-diversity while for larger values they may expect above $0.06$-diversity. We can understand these values of diversity as a proportion of the amount injected. A value of $\delta = 0.06$ means the proportion of tweets seen of each type is at least 6\% of the magnitude of injected tweets. Although this may seem low, as there are four types, so in total 24\% of the magnitude is necessarily accounted for by simply showing diverse content. If the policy-maker is willing to sacrifice 10\%, this number shoots up to about $0.07$-diversity for lower scales and above $0.10$ for higher, i.e., over 40\% of the magnitude already accounted for.

Finally, another interesting observation is that increasing the scale very reliably improves the tradeoff. However, strikingly, there seem to exist exceptions, namely that with the mode probabilities, the probabilities scaled up by 30 perform worse than those scaled up by 10 for large levels of diversity. This demonstrates a surprisingly nuanced relation between the magnitude of probabilities and the diversity-engagement tradeoff, which may be difficult to characterize analytically. 

\section{Discussion}\label{sec:disc}

While our model may appear stylized, we believe that it is quite robust. In essence, the main assumptions are a partition of tweets into types and known retweet probabilities, both of which seem quite reasonable. 

Some issues that are ostensibly outside the scope of our model can, in fact, be captured by it. One is that not all tweets that fall into even a specific type, such as ``climate change,'' have the same retweet probabilities. In theory, one could make the set of types arbitrarily granular, but this would make diversity constraints impractical. A better approach, which we believe to be plausible, is to set $p_{ti}$ (the probability of user $i$ retweeting type $t$) to be the \emph{average} of the retweet probabilities of user $i$ for different tweets that are included in type $t$. 

Another seemingly restrictive modeling choice that can easily be relaxed is the fact that a user's engagement is defined with respect to their retweet probability rather than a distinct ``type engagement'' parameter (or ``type affinity''). Differentiating these features would allow the model to capture users that perhaps have high engagement yet rarely retweet. We could have instead introduced an additional parameter $e_{ti}$ for each user and type to be used in the definition of engagement, i.e., $\eng(\x) = \sum_t \inner{\e_t, \x_t}$. At a technical level, this hardly seems to affect the model. In the linear programs described in \Cref{subsec:optimizing}, the $\p_t$ vectors in the objective would be replaced with $\e_t$. It can be checked that the proof of \Cref{thm:app-optimal} goes through, and even \Cref{thm:main} does as well with the additional condition that the average and maximum affinities also satisfy the $\alpha$ and $\beta$ constraints. Our choice to exclude it was solely for presentation, as we did not believe the gain in generality was worth the loss in comprehensibility in a paper already defining half the alphabet.

That said, we readily acknowledge that our model has limitations. To name one, we view propagation dynamics in the social network as resulting from retweets of content that is injected by the platform. But users also create content; for example, a political reporter will likely write new tweets about politics. This can be modeled as another injection policy that is outside of our control, but it is unclear what values one would choose for this policy. 

Nevertheless, in our view, our model and analysis provide useful insights into the engagement-diversity tradeoff. As discussed in \Cref{sec:intro}, however, the jury is still out on the diversity-polarization connection, and it is a topic of intensive inquiry. With a better (quantitative) understanding of this connection, our results could be directly leveraged to analyze engagement-polarization tradeoffs, potentially helping social media platforms curb negative societal impacts.

\bibliographystyle{ACM-Reference-Format}
\bibliography{abb,twitter}


\begin{thebibliography}{24}


\ifx \showCODEN    \undefined \def \showCODEN     #1{\unskip}     \fi
\ifx \showDOI      \undefined \def \showDOI       #1{#1}\fi
\ifx \showISBNx    \undefined \def \showISBNx     #1{\unskip}     \fi
\ifx \showISBNxiii \undefined \def \showISBNxiii  #1{\unskip}     \fi
\ifx \showISSN     \undefined \def \showISSN      #1{\unskip}     \fi
\ifx \showLCCN     \undefined \def \showLCCN      #1{\unskip}     \fi
\ifx \shownote     \undefined \def \shownote      #1{#1}          \fi
\ifx \showarticletitle \undefined \def \showarticletitle #1{#1}   \fi
\ifx \showURL      \undefined \def \showURL       {\relax}        \fi
\providecommand\bibfield[2]{#2}
\providecommand\bibinfo[2]{#2}
\providecommand\natexlab[1]{#1}
\providecommand\showeprint[2][]{arXiv:#2}

\bibitem[Amati et~al\mbox{.}(2021)]%
        {amati2021topic}
\bibfield{author}{\bibinfo{person}{G. Amati}, \bibinfo{person}{S. Angelini},
  \bibinfo{person}{A. Cruciani}, \bibinfo{person}{G. Fusco},
  \bibinfo{person}{G. Gaudino}, \bibinfo{person}{D. Pasquini}, {and}
  \bibinfo{person}{P. Vocca}.} \bibinfo{year}{2021}\natexlab{}.
\newblock \showarticletitle{Topic Modeling by Community Detection Algorithms}.
  In \bibinfo{booktitle}{\emph{Proceedings of the 1st Workshop on Open
  Challenges in Online Social Networks (OASIS)}}. \bibinfo{pages}{15--20}.
\newblock


\bibitem[Anderson et~al\mbox{.}(2020)]%
        {AMMA+20}
\bibfield{author}{\bibinfo{person}{A. Anderson}, \bibinfo{person}{L. Maystre},
  \bibinfo{person}{R. Mehrotra}, \bibinfo{person}{I. Anderson}, {and}
  \bibinfo{person}{M. Lalmas}.} \bibinfo{year}{2020}\natexlab{}.
\newblock \showarticletitle{Algorithmic Effects on the Diversity of Consumption
  on Spotify}. In \bibinfo{booktitle}{\emph{Proceedings of the 29th
  International World Wide Web Conference (WWW)}}. \bibinfo{pages}{2155--2165}.
\newblock


\bibitem[Bail(2018)]%
        {Bail18}
\bibfield{author}{\bibinfo{person}{C.~A. Bail}.}
  \bibinfo{year}{2018}\natexlab{}.
\newblock \bibinfo{title}{Please, {T}witter, Don't Do It}.
\newblock \bibinfo{howpublished}{The New York Times, September 9}.
\newblock


\bibitem[Bail et~al\mbox{.}(2018)]%
        {BABB+18}
\bibfield{author}{\bibinfo{person}{C.~A. Bail}, \bibinfo{person}{L.~P. Argyle},
  \bibinfo{person}{T.~W. Brown}, \bibinfo{person}{J.~P. Bumpus},
  \bibinfo{person}{H. Chen}, \bibinfo{person}{M.~B.~F. Hunzaker},
  \bibinfo{person}{J. Lee}, \bibinfo{person}{M. Mann}, \bibinfo{person}{F.
  Merhout}, {and} \bibinfo{person}{A. Volfovsky}.}
  \bibinfo{year}{2018}\natexlab{}.
\newblock \showarticletitle{Exposure to Opposing Views on Social Media Can
  Increase Political Polarization}.
\newblock \bibinfo{journal}{\emph{Proceedings of the National Academy of
  Sciences}} \bibinfo{volume}{115}, \bibinfo{number}{37}
  (\bibinfo{year}{2018}), \bibinfo{pages}{9216--9221}.
\newblock


\bibitem[Bakshy et~al\mbox{.}(2015)]%
        {BMA15}
\bibfield{author}{\bibinfo{person}{E. Bakshy}, \bibinfo{person}{S. Messing},
  {and} \bibinfo{person}{L.~A. Adamic}.} \bibinfo{year}{2015}\natexlab{}.
\newblock \showarticletitle{Exposure to Ideologically Diverse News and Opinion
  on {F}acebook}.
\newblock \bibinfo{journal}{\emph{Science}} \bibinfo{volume}{348},
  \bibinfo{number}{6239} (\bibinfo{year}{2015}), \bibinfo{pages}{1130--1132}.
\newblock


\bibitem[Blondel et~al\mbox{.}(2008)]%
        {blondel2008fast}
\bibfield{author}{\bibinfo{person}{V.~D. Blondel}, \bibinfo{person}{J.-L.
  Guillaume}, \bibinfo{person}{R. Lambiotte}, {and} \bibinfo{person}{E.
  Lefebvre}.} \bibinfo{year}{2008}\natexlab{}.
\newblock \showarticletitle{Fast Unfolding of Communities in Large Networks}.
\newblock \bibinfo{journal}{\emph{Journal of Statistical Mechanics: Theory and
  Experiment}} \bibinfo{volume}{2008}, \bibinfo{number}{10}
  (\bibinfo{year}{2008}), \bibinfo{pages}{P10008}.
\newblock


\bibitem[Conover et~al\mbox{.}(2011)]%
        {CRFG+11}
\bibfield{author}{\bibinfo{person}{M. Conover}, \bibinfo{person}{J.
  Ratkiewicz}, \bibinfo{person}{M. Francisco}, \bibinfo{person}{B. Goncalves},
  \bibinfo{person}{F. Menczer}, {and} \bibinfo{person}{A. Flammini}.}
  \bibinfo{year}{2011}\natexlab{}.
\newblock \showarticletitle{Political Polarization on {T}witter}. In
  \bibinfo{booktitle}{\emph{Proceedings of the 5th International AAAI
  Conference on Web and Social Media (ICWSM)}}. \bibinfo{pages}{89--96}.
\newblock


\bibitem[Flaxman et~al\mbox{.}(2016)]%
        {FGR16}
\bibfield{author}{\bibinfo{person}{S. Flaxman}, \bibinfo{person}{S. Goel},
  {and} \bibinfo{person}{J.~M. Rao}.} \bibinfo{year}{2016}\natexlab{}.
\newblock \showarticletitle{Filter Bubbles, Echo Chambers, and Online News
  Consumption}.
\newblock \bibinfo{journal}{\emph{Public Opinion Quarterly}}
  \bibinfo{volume}{80} (\bibinfo{year}{2016}), \bibinfo{pages}{298--320}.
\newblock


\bibitem[Garimella et~al\mbox{.}(2018a)]%
        {garimella1}
\bibfield{author}{\bibinfo{person}{V.~R.~K. Garimella}, \bibinfo{person}{G. {De
  Francisci Morales}}, \bibinfo{person}{A. Gionis}, {and} \bibinfo{person}{M.
  Mathioudakis}.} \bibinfo{year}{2018}\natexlab{a}.
\newblock \showarticletitle{Political Discourse on Social Media: Echo Chambers,
  Gatekeepers, and the Price of Bipartisanship}. In
  \bibinfo{booktitle}{\emph{Proceedings of the 27th International World Wide
  Web Conference (WWW)}}. \bibinfo{pages}{913--922}.
\newblock


\bibitem[Garimella et~al\mbox{.}(2018b)]%
        {garimella3}
\bibfield{author}{\bibinfo{person}{V.~R.~K. Garimella}, \bibinfo{person}{G. {De
  Francisci Morales}}, \bibinfo{person}{A. Gionis}, {and} \bibinfo{person}{M.
  Mathioudakis}.} \bibinfo{year}{2018}\natexlab{b}.
\newblock \showarticletitle{Quantifying Controversy on Social Media}.
\newblock \bibinfo{journal}{\emph{ACM Transactions on Social Computing}}
  \bibinfo{volume}{1}, \bibinfo{number}{1} (\bibinfo{year}{2018}),
  \bibinfo{pages}{3:1--3:27}.
\newblock


\bibitem[Garimella and Weber(2017)]%
        {garimella2}
\bibfield{author}{\bibinfo{person}{V.~R.~K. Garimella} {and}
  \bibinfo{person}{I. Weber}.} \bibinfo{year}{2017}\natexlab{}.
\newblock \showarticletitle{A Long-Term Analysis of Polarization on Twitter}.
  In \bibinfo{booktitle}{\emph{Proceedings of the 11th International AAAI
  Conference on Web and Social Media (ICWSM)}}. \bibinfo{pages}{528--531}.
\newblock


\bibitem[Holtz et~al\mbox{.}(2020)]%
        {HCCN+20}
\bibfield{author}{\bibinfo{person}{D. Holtz}, \bibinfo{person}{B. Carterette},
  \bibinfo{person}{P. Chandar}, \bibinfo{person}{Z. Nazari},
  \bibinfo{person}{H. Cramer}, {and} \bibinfo{person}{S. Aral}.}
  \bibinfo{year}{2020}\natexlab{}.
\newblock \showarticletitle{The Engagement-Diversity Connection: Evidence from
  a Field Experiment on Spotify}. In \bibinfo{booktitle}{\emph{Proceedings of
  the 21st ACM Conference on Economics and Computation (EC)}}.
  \bibinfo{pages}{75--76}.
\newblock


\bibitem[Hubbard and Hubbard(2015)]%
        {hubbard2015vector}
\bibfield{author}{\bibinfo{person}{J.~H. Hubbard} {and} \bibinfo{person}{B.~B.
  Hubbard}.} \bibinfo{year}{2015}\natexlab{}.
\newblock \bibinfo{booktitle}{\emph{Vector Calculus, Linear Algebra, and
  Differential Forms: A Unified Approach}}.
\newblock \bibinfo{publisher}{Matrix Editions}.
\newblock


\bibitem[Husz\'ar et~al\mbox{.}(2021)]%
        {HKOB+21}
\bibfield{author}{\bibinfo{person}{F. Husz\'ar}, \bibinfo{person}{S.~I. Ktena},
  \bibinfo{person}{C. O'Brien}, \bibinfo{person}{L. Belli}, \bibinfo{person}{A.
  Schlaikjer}, {and} \bibinfo{person}{M. Hardt}.}
  \bibinfo{year}{2021}\natexlab{}.
\newblock \showarticletitle{Algorithmic Amplification of Politics on Twitter}.
\newblock \bibinfo{journal}{\emph{Proceedings of the National Academy of
  Sciences}} \bibinfo{volume}{119}, \bibinfo{number}{1} (\bibinfo{year}{2021}),
  \bibinfo{pages}{e2025334119}.
\newblock


\bibitem[Johnson et~al\mbox{.}(2020)]%
        {JKG20}
\bibfield{author}{\bibinfo{person}{S.~L. Johnson}, \bibinfo{person}{B.
  Kitchens}, {and} \bibinfo{person}{P. Gray}.} \bibinfo{year}{2020}\natexlab{}.
\newblock \bibinfo{title}{Facebook Serves as an Echo Chamber, Especially for
  Conservatives. {B}lame its algorithm.}
\newblock \bibinfo{howpublished}{The Washington Post, October 26}.
\newblock


\bibitem[Kitchens et~al\mbox{.}(2020)]%
        {KJG20}
\bibfield{author}{\bibinfo{person}{B. Kitchens}, \bibinfo{person}{S.~L.
  Johnson}, {and} \bibinfo{person}{P. Gray}.} \bibinfo{year}{2020}\natexlab{}.
\newblock \showarticletitle{Understanding Echo Chambers and Filter Bubbles: The
  Impact of Social Media on Diversification and Partisan Shifts in News
  Consumption}.
\newblock \bibinfo{journal}{\emph{MIS Quarterly}} \bibinfo{volume}{44},
  \bibinfo{number}{4} (\bibinfo{year}{2020}), \bibinfo{pages}{1619--1649}.
\newblock


\bibitem[Lanier(2022)]%
        {Lan22}
\bibfield{author}{\bibinfo{person}{J. Lanier}.}
  \bibinfo{year}{2022}\natexlab{}.
\newblock \bibinfo{title}{{T}rump, {M}usk and {K}anye Are {T}witter Poisoned}.
\newblock \bibinfo{howpublished}{The New York Times, November 13}.
\newblock


\bibitem[Lee and Hosanagar(2019)]%
        {LH19}
\bibfield{author}{\bibinfo{person}{D. Lee} {and} \bibinfo{person}{K.
  Hosanagar}.} \bibinfo{year}{2019}\natexlab{}.
\newblock \showarticletitle{How Do Recommender Systems Affect Sales Diversity?
  A Cross-Category Investigation via Randomized Field Experiment}.
\newblock \bibinfo{journal}{\emph{Information Systems Research}}
  \bibinfo{volume}{30}, \bibinfo{number}{1} (\bibinfo{year}{2019}),
  \bibinfo{pages}{239--259}.
\newblock


\bibitem[Pariser(2011)]%
        {Par11}
\bibfield{author}{\bibinfo{person}{E. Pariser}.}
  \bibinfo{year}{2011}\natexlab{}.
\newblock \bibinfo{booktitle}{\emph{The Filter Bubble: How the New Personalized
  Web Is Changing What We Read and How We Think}}.
\newblock \bibinfo{publisher}{Penguin}.
\newblock


\bibitem[Pervin et~al\mbox{.}(2015)]%
        {pervin2015hashtag}
\bibfield{author}{\bibinfo{person}{N. Pervin}, \bibinfo{person}{T.~Q. Phan},
  \bibinfo{person}{A. Datta}, \bibinfo{person}{H. Takeda}, {and}
  \bibinfo{person}{F. Toriumi}.} \bibinfo{year}{2015}\natexlab{}.
\newblock \showarticletitle{Hashtag Popularity on Twitter: Analyzing
  Co-Occurrence of Multiple Hashtags}. In \bibinfo{booktitle}{\emph{Proceedings
  of the 7th International Conference on Social Computing and Social Media
  (SCSM)}}. \bibinfo{pages}{169--182}.
\newblock


\bibitem[Rudin(1991)]%
        {rudin1991functional}
\bibfield{author}{\bibinfo{person}{W. Rudin}.} \bibinfo{year}{1991}\natexlab{}.
\newblock \bibinfo{booktitle}{\emph{Functional Analysis}
  (\bibinfo{edition}{2nd} ed.)}.
\newblock \bibinfo{publisher}{McGraw-Hill}.
\newblock


\bibitem[Saveski et~al\mbox{.}(2022)]%
        {SBMR22}
\bibfield{author}{\bibinfo{person}{M. Saveski}, \bibinfo{person}{D. Beeferman},
  \bibinfo{person}{D. McClure}, {and} \bibinfo{person}{D. Roy}.}
  \bibinfo{year}{2022}\natexlab{}.
\newblock \showarticletitle{Engaging Politically Diverse Audiences on Social
  Media}. In \bibinfo{booktitle}{\emph{Proceedings of the 16th International
  AAAI Conference on Web and Social Media (ICWSM)}}. \bibinfo{pages}{873--884}.
\newblock


\bibitem[Stroud(2010)]%
        {Stroud10}
\bibfield{author}{\bibinfo{person}{N.~J. Stroud}.}
  \bibinfo{year}{2010}\natexlab{}.
\newblock \showarticletitle{Polarization and Partisan Selective Exposure}.
\newblock \bibinfo{journal}{\emph{Journal of Communication}}
  \bibinfo{volume}{60}, \bibinfo{number}{3} (\bibinfo{year}{2010}),
  \bibinfo{pages}{556--576}.
\newblock


\bibitem[Sunstein(2018)]%
        {Suns18}
\bibfield{author}{\bibinfo{person}{C.~R. Sunstein}.}
  \bibinfo{year}{2018}\natexlab{}.
\newblock \bibinfo{booktitle}{\emph{\#Republic: Divided Democracy in the Age of
  Social Media}}.
\newblock \bibinfo{publisher}{Princeton University Press}.
\newblock


\end{thebibliography}
\newpage
\appendix

\end{document}